\documentclass[journal,10 pt]{IEEEtran}
\IEEEoverridecommandlockouts                            

\usepackage{color}
\usepackage{enumerate,graphicx,epstopdf}
\usepackage{amsmath,amssymb,bm}
\usepackage{cite}
\usepackage{mathtools}
\usepackage{array}
\usepackage{algpseudocode,algorithm,algorithmicx}
\usepackage{booktabs}
\usepackage[hyphens]{url}
\usepackage{lipsum}
 \usepackage[dvipsnames]{xcolor}
\usepackage[deletedmarkup=sout,authormarkup=superscript]{changes}
\usepackage[normalem]{ulem}
\usepackage{flushend}
\usepackage[hidelinks]{hyperref}
\usepackage{enumitem}
\usepackage{amsmath}
\usepackage{subcaption}
\usepackage{booktabs}

\DeclareMathOperator*{\argmin}{arg\,min}

\definechangesauthor[name={Mamzi}, color=NavyBlue]{MA}
\definechangesauthor[name={Dominic}, color=RedViolet]{DLM}
\definechangesauthor[name={Efe}, color=OliveGreen]{EB}

\newcommand{\mc}[0] {\mathcal}
\newtheorem{remark}{Remark}

\title{
Layer-to-Layer Melt Pool Control in \\ Laser Powder Bed Fusion
}

\author{Dominic Liao-McPherson$^{*}$, Efe C. Balta$^*$, Mohamadreza Afrasiabi$^*$,\\ Alisa Rupenyan, Markus Bambach, and John Lygeros
\thanks{$^{*}$These authors contributed equally to this work.\\ 
D. Liao-McPherson is with the University of British Columbia, Vancouver, Canada. \texttt{dliaomcp@mech.ubc.ca} \\%
E. C. Balta is with the inspire AG Control and Automation Group and the ETH Zürich Automatic Control Laboratory, Zürich, Switzerland. \texttt{efe.balta@inspire.ch}. \\
M. Afrasiabi is with the inspire AG Computational Manufacturing Group and ETH Zürich Advanced Manufacturing Laboratory,  Zürich, Switzerland. \texttt{mamzi.afrasiabi@inspire.ch} \\%
A. Rupenyan is with the ZHAW Centre for Artificial Intelligence, ZHAW Z\"{u}rich University of Applied Sciences, Zürich, Switzerland. \texttt{rupn@zhaw.ch} \\%
M. Bambach is with the ETH Zürich Advanced Manufacturing Laboratory, \texttt{mbambach@ethz.ch}\\%
J. Lygeros is with the ETH Zürich Automatic Control Laboratory (IfA), \texttt{ jlygeros@ethz.ch}\\%
This work is supported by the Swiss National Science Foundation via NCCR Automation (Grant Number 180545)
}
}

\begin{document}

\maketitle

\begin{abstract}
Additive manufacturing processes are flexible and efficient technologies for producing complex geometries. However, ensuring reliability and repeatability is challenging due to the complex physics and various sources of uncertainty in the process. In this work, we investigate closed-loop control of the melt pool dimensions in 2D laser powder bed fusion (LPBF) process. We propose a trajectory optimization-based layer-to-layer controller based on a linear parameter varying model that adjusts the laser power input to the next layer to track a desired melt pool depth and validate our controller by placing it in closed-loop high-fidelity multi-layer smoothed particle hydrodynamics simulator of the 2D LPBF process. Detailed numerical case studies demonstrate successful regulation of the melt pool depth on brick and overhang geometries and provide first of its kind results on the effectiveness of layer-to-layer input optimization for the LPBF process as well as detailed insight into the physics of the controlled process. Computational complexity and process performance results illustrate the method's effectiveness and provide an outlook for its implementation onto real systems.\\
\end{abstract}

\section{Introduction}
\label{sec:intro}
\begin{figure}
    \centering
    \includegraphics[width=0.4\textwidth]{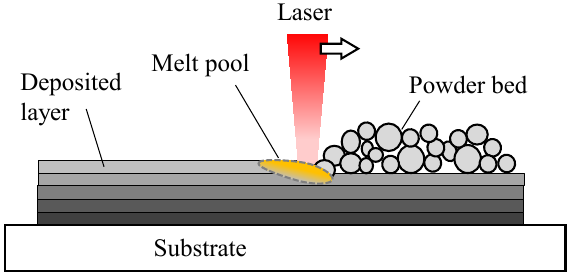}
    \caption{In an LPBF process, a thin layer of metal powder is selectively melted using a high-power laser. The melted powder material then solidifies and bonds to the underlying solid volume.}
    \label{fig:slm_schematic}
\end{figure}

Additive manufacturing (AM) is a family of production technologies that build up three-dimensional (3D) objects layer-by-layer. AM processes are flexible and can produce intricate product geometries using a variety of materials. Among different AM technologies, laser powder bed fusion (LPBF), illustrated in Figure~\ref{fig:slm_schematic}, is the most widely-used process for metals due to its versatility and maturity. LPBF's ability to create highly complex parts, its scalability and short lead times make it an attractive option for a wide range of applications in various industries.

One of the key challenges in using LPBF and similar AM processes, is ensuring reliability and repeatability under changing process conditions and for different part geometries~\cite{dowling2020review}. Ensuring repeatability is challenging due to the complex physics involved in the process (e.g., material phase transitions), as well as external disturbances. Parameter optimization has been widely studied to remedy these issues~\cite{ahmed2022process}. However, parameter optimization cannot effectively account for run-time uncertainties and disturbances, which are often the cause of processes variability. One method for addressing the aforementioned issues in AM processes is closed-loop control~\cite{mani2017review,tapia2014review}, which has been proposed using process measurements at inter-layer and intra-layer timescales to optimize process performance and account for disturbances and uncertainties at run-time for various metal AM processes including LPBF~\cite{benda1994temperature,kruth2007feedback,shkoruta2022real,kavas2023layer}, wire arc AM~\cite{heralic2012height,hagqvist2015resistance,xia2020review,mu2022layer}, and metal deposition~\cite{sammons2013height,sammons2018repetitive}.

The shape, size, and thermal profile of the melt pool are the primary factors affecting the properties of the final part. They govern the solidification behavior and cooling rates, which in turn impact the microstructure and porosity of the printed component \cite{debroy2018additive,
bajaj2020steels}. The melt pool should be sufficiently deep to ensure strong bonding to the layer below while avoiding excessive heat transfer into previously printed layers to reduce thermal stresses in the final part.

The two main approaches for LPBF control have been proposed in the literature: In-layer (intra-layer) controllers update process inputs during the printing of a single layer, while inter-layer or layer-to-layer (L2L) controllers update the inputs between the layers.
In-layer control has been used to adjust melt pool dimensions in experimental setups on single tracks and build parts using PID control~\cite{benda1994temperature,renken2019process,Renken2018,shkoruta2022real,wang2020model}, on geometries with overhang features~\cite{kruth2007feedback}, and with data-driven modeling techniques in simulation studies~\cite{asadi2021gaussian}.
While providing promising results, these methods rely on controller tuning to get desired process characteristics, which may be time consuming and expensive. Additionally, the tuning is often related to print geometry and other process parameters, and may need to be re-calibrated for each new part.
A major technical challenge for in-layer control architectures is the high sampling rate of the in-layer dynamics, which requires ultra-fast sensing, actuation, and computing. 

Layer-to-layer (L2L) controllers update the input signal for the next layer based on measurements gathered during the printing of previous layers. This results in a slower update rate, enabling the optimization of more complex control objectives and avoiding the need for ultra-fast sensors and actuators. An L2L control-oriented model for LBPF is presented in~\cite{wang2020layer}. Iterative learning control for updating input power in an LPBF process to control surface temperature is investigated in \cite{spector2018passivity} but uses only a simplified ordinary differential equation (ODE)-based single-layer model. Model predictive control applied to the same ODE model is able to effectively control surface temperature~\cite{zuliani2022batch}. A multi-layer extension considering multiple layers based on the linear quadratic regulator is proposed in~\cite{liao2022layer} while a model-free iterative learning control approach for temperature regulation is proposed in \cite{inyang2022model} and validated using a high-fidelity single-layer simulator. Finally, an adaptive proportional controller is tested experimentally in \cite{rezaeifar2021line} to control the melt pool temperature at a single instant during a layered print based on pyrometer measurements.

There is relatively little literature on L2L regulation of the melt pool dimensions in LPBF due to the difficulty of simulating the melt pool in multi-layer processes and the complexity of monitoring melt pool behavior on experimental systems. A recent study~\cite{vasileska2022novel} investigates L2L control of laser power based on melt pool images on an experimental setup without considering input constraints or optimization while \cite{park2022melt} proposes an iterative learning controller for the melt pool area based on camera images.
Furthermore, while experimental results show promising improvements, a detailed analysis of process characteristics when using closed-loop control is missing in the current literature due to a lack of efficient numerical simulations and models for multi-layer LPBF processes. 
Studying the process parameters in closed-loop and understanding efficient control architectures and formulations is critical as such studies enable the next level of high-performance LPBF processes.

This work proposes a novel L2L controller for LPBF that uses trajectory optimization-based control to regulate the dimensions of the melt pool by adjusting the laser power input based on end-of-layer measurements. We validate our proposed controller in closed-loop with a high-fidelity multi-layer simulation model and demonstrate its ability to regulate the melt pool depth on block and overhang geometries by automatically compensating for heat/energy build-up in the part. In this paper, we focus on laser power and leave adjusting other process parameters such as the laser speed or the scan track for future work.

Our method makes use of a high-fidelity characterization of the process physics and its interplay with changing process parameters, which is invaluable for designing high-performance L2L controllers and obtaining detailed insights into the physics of the closed-loop process. High-fidelity simulation of LPBF processes at the powder scale (i.e., mesoscale) is a daunting but essential task to establish a deep understanding of the process for control and optimization purposes. This has been highlighted in previous works such as \cite{king2015laser,bayat2021review,mehrpouya2022multimaterial}. 

Simulating LPBF processes at the mesoscale introduces various computational and material challenges across different time and length scales. These challenges primarily arise from the highly dynamic nature of multi-phase flows, characterized by violent free-surface movements, high solidification rates, and rapid phase changes. Detailed discussions on these challenges can be found in \cite{francois2017modeling,yan2018modeling}. Several studies have focused on addressing these computational challenges by developing mesh-based computational fluid dynamics (CFD) \cite{khairallah2016laser,yu2022quantitative} and particle-based modeling frameworks \cite{furstenau2020generating,afrasiabi2021multi,meier2021novel} for high-fidelity melt pool simulations. However, none of these studies have utilized the generated numerical results for LPBF process control, which is the ultimate goal of developing such simulations in the first place. While the utilization of detailed and high-resolution 3D simulations for process control in LPBF is currently not feasible due to their extensive computational times, it is essential to explore the use of 2D models as an initial step toward deriving a process controller. This approach may entail some loss of accuracy in modeling certain phenomena that cannot be fully represented by 2D models (e.g., heat transfer), but it remains the most reasonable strategy for the initial investigation.

\begin{figure}[htbp]
    \centering
    \includegraphics[width=\columnwidth]{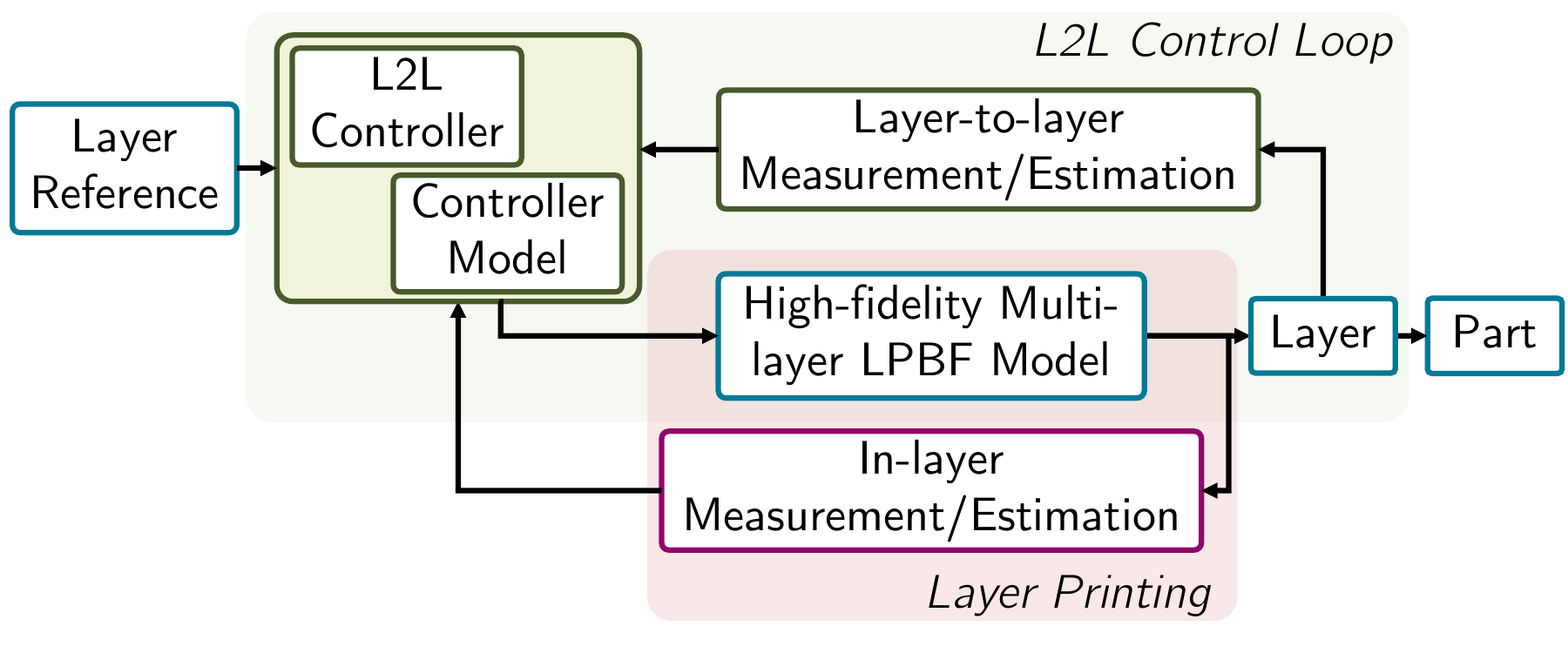}
    \caption{LPBF closed-loop control architecture considered in this work.}
    \label{fig:overall-arch}
\end{figure}

Our proposed control architecture is summarized in Fig.~\ref{fig:overall-arch}. We use the high-fidelity simulator as our plant and propose a controller that takes surface temperature measurements at the end of the layer and computes a laser power input signal for the next layer.

This work presents the first closed-loop high-fidelity multi-layer simulation of an L2L-controlled LPBF process. Our objective is to investigate the feasibility and potential impact of controlling the melt pool dimensions in an LPBF process. Our controller can incorporate optimization objectives and enforce input constraints which is not possible using the strategies proposed in \cite{vasileska2022novel,park2022melt}. 
In this work we use a linear parameter varying surrogate process model which works well in 2D and our methodology is compatible with the more expressive models needed for complex 3D geometries. Additionally, the high-fidelity simulator provides deeper insights into how control, sensing, and actuation interact with and impact the physics of the process than would be possible with experiments. We study parts with continuous layers and present detailed simulation results about a number of process parameters that are not in-situ measurable on a real process, building a key understanding of the closed-loop process behavior of LPBF. 

 This understanding represents a critical first step towards experimental implementation and serves as a preliminary ``proof-of-concept'' to motivate further development.

Section~\ref{sec:method} introduces the high-fidelity multi-layer process simulation. Section~\ref{sec:control} provides details of the proposed L2L control strategy including the underlying optimization problem and the data-driven prediction models used internally by the controller. Section~\ref{sec:res} presents our detailed closed-loop simulation studies on two representative print geometries. Finally, an outlook for implementation on real systems is provided in Section~\ref{sec:outlook}.

\section{High-fidelity Multi-layer LPBF Process Model} \label{sec:method}
To model the multi-layer LPBF process with a high degree of fidelity, our numerical simulation framework needs to solve the coupled thermo-fluid dynamics equations (including phase transitions) to capture the dynamic behaviour of the melt pool and model the recoating step (powder deposition and cooling between layers). We use an efficient meshfree modeling approach based on the SPH (smoothed particle hydrodynamics) method, developed by Afrasiabi et al. \cite{afrasiabi2022smoothed}. We make the following assumptions:
\begin{enumerate}[noitemsep]
    \item The simulation domain is 2D
    \item The heat source is volumetric and obeys the Beer-Lambert law
    \item The laser moves with a constant speed
    \item The flow of the molten material is incompressible
    \item Gas phase and evaporation are not considered
    \item All powder grains are spherical (i.e., circular in 2D)
\end{enumerate}
A brief description of the process simulator's main \emph{melt pool simulation} and \emph{powder layer generation} modules is presented next. A detailed description of the predictive capabilities and experimental validation of our SPH-based LPBF model for 2D and 3D single-track applications is available in \cite{afrasiabi2021multi,afrasiabi2022smoothed,afrasiabi2022effect}. 

Note that for control design we limit our attention to 2D simulations due to the prohibitively high computational costs associated with 3D simulations. The efficient codes developed for powder-scale LPBF simulations, employing hundreds or thousands of CPU/GPU cores \cite{furstenau2020generating,dao2021simulations}, or adaptive resolutions \cite{luethi2023adaptive}, can take several hours to a few days to model a single powder layer of approximately 1 mm. This makes it impractical to perform complete 3D simulations of multi-layer LPBF, even with the computing power available today. To generate sufficient training data for the layer-to-layer controller, we perform a series of 2D LPBF simulations. These simulations can model a 2-mm layer within minutes and are better suited for generating large datasets.

\subsection{Powder Layer Generation Module}
\label{subsec:HF_powder}
The module implements a so-called \textit{rainfall} approach, whereby the powder grains generated above the substrate fall on the previous (solidified) track under their gravitational body force. We consider a penalty-based spring dash-pot model without frictional forces to account for solid-solid interactions between the powder grains. Each powder grain $k$ in this framework is treated as one discrete element represented by several SPH particles, on which the resultant force $\vec{F}_{k}$ and torque $\vec{\tau}_{k}$ are exerted:
\begin{equation}
\displaystyle \vec{F}_{k}  = \sum_{i\in k} \vec{f}_{i}, \quad \vec{\tau}_{k}  = \sum_{i\in k} \vec{r}_{iM} \times \vec{f}_{i}
\label{eq:laws_of_motion}
\end{equation}
where $\vec{f}_{i}$ denotes the collision force contribution of particle $i$ computed from the non-linear Hertzian elastic contact formulation (including the repulsive and damping components) and $\vec{r}_{iM}$ is the distance vector to the grain's center of mass. See \cite{canelas2013generalized,afrasiabi2022smoothed} for more detailed derivation and implementation procedures.

The powder generation module at each layer is run over a time interval of 1s, ensuring sufficient time for the powder particles to settle down. Powder grains exiting the substrate's bounding walls are subsequently deleted from the configuration in order to prevent numerical instability during the melt pool simulation. More details regarding the layer thickness and cooling time between the layers are provided in Section \ref{sec:sim_setup}.

\subsection{Melt Pool Simulation Module}
\label{subsec:HF_lpbf}
The power layer generation module is coupled with a thermal-fluid dynamics solver to render a complete picture of the multi-layer LPBF process. Given our incompressible flow assumption, behaviour of the melt pool is governed by mass, momentum, and energy conservation expressed by the Navier-Stokes and heat transfer equations. In a Lagrangian frame, these balance equations are:
\begin{align}
\dot{\rho} & = -\rho\nabla\cdot\vec{v} \label{eq:contin} \\ 
\rho\;\dot{\vec{v}} & = -\nabla p + \mu \nabla^2\vec{v} + \vec{f}_s + \rho \vec{g} \label{eq:moment} \\
\rho\,c_{p}\;\dot{T} & = \bm{\tau}:\nabla\vec{v}+\nabla\cdot(k \nabla T)
\label{eq:heat_cond}
\end{align}
\sloppy where $\rho$ is the density, $\vec{v}$ the velocity, $p$ the pressure, $\mu$ the viscosity, $\vec{f}_s$ the force contribution due to surface tension and Marangoni effects, $\vec{g}$ the gravitational acceleration, $c_p$ the specific heat capacity, $T$ the temperature, $\bm{\tau}$ the shear stress tensor, $k$ the thermal conductivity, and $:$ denotes the double inner product. Within this numerical process model, the apparent latent heat associated with the phase change is considered through a modified temperature-dependent heat capacity coefficient, as described in our previous work \cite{afrasiabi2021multi}. The laser input energy is included in the thermal boundary conditions associated with \eqref{eq:heat_cond} along with the overall heat loss through the open surfaces (i.e., radiation and convection). Specifically, the last input is modeled in our SPH code as a volumetric heat source based on the Beer-Lambert law with a normalized Gaussian distribution for the laser beam intensity, computed from:
\begin{equation}
I(r)=\frac{2u}{\pi R^2}\exp\left(-2\left(\frac{r}{R}\right)^2\right)
\label{eq:gaussian_beam}
\end{equation}
in which $u$ is the laser power (which is the main control input for this process) and $R$ is the laser beam radius. The governing PDEs expressed in Eqs. \eqref{eq:contin}-\eqref{eq:heat_cond} are discretized in space by the SPH approximation operators and then integrated in time by a second-order predictor-corrector method. 
\begin{figure}[h]
    \centering
    \includegraphics[width=\columnwidth]{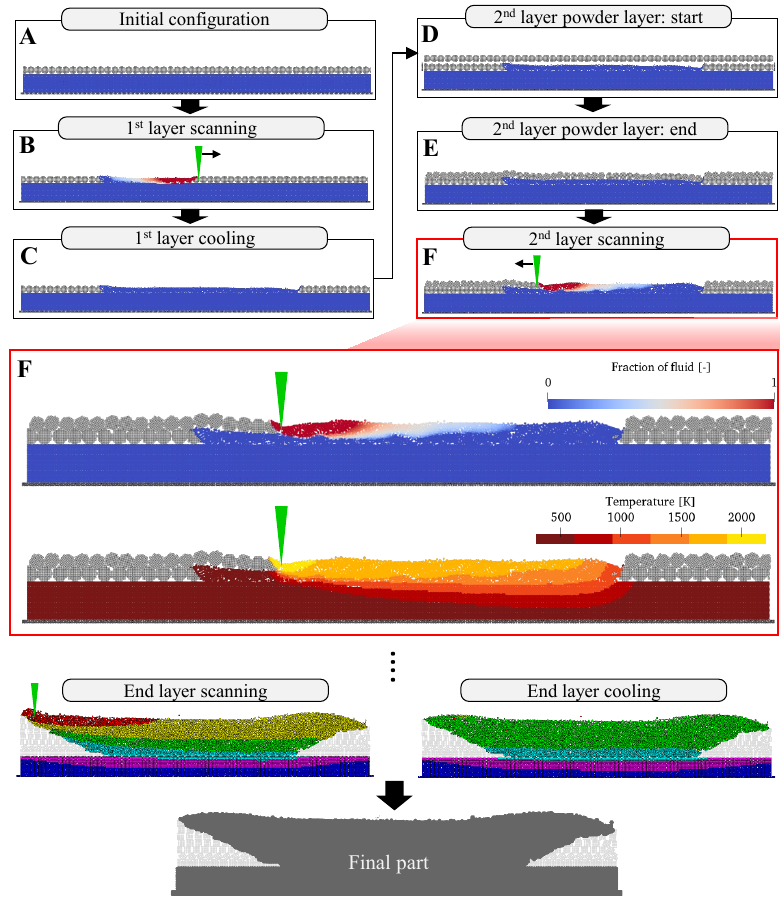}
    \caption{Consecutive steps of the multi-layer LPBF simulation in our SPH-based modeling framework: (A)-(F). Fraction of fluid and temperature distributions at layer 2 are shown for a time-step in which the melt pool has reached a steady state. {The color spectrum from red to blue in the two images below the subfigure \textbf{F} depicts the transition from hot to cold temperatures.}}
    \label{fig:sim_full_steps}
\end{figure}

The simulation process is illustrated in Figure~\ref{fig:sim_full_steps}. The simulation proceeds layer by layer and alternates between powder generation, laser scanning, and cooling phases. The control input is the laser power profile $u(t)$ for each layer. At the end of each layer, our proposed L2L controller uses the surface temperature distribution extracted from the simulation to compute a laser power input profile for the next layer.

\section{Control Design}
\label{sec:control}
\begin{figure}[ht]
    \centering
    \includegraphics[width=0.9\columnwidth]{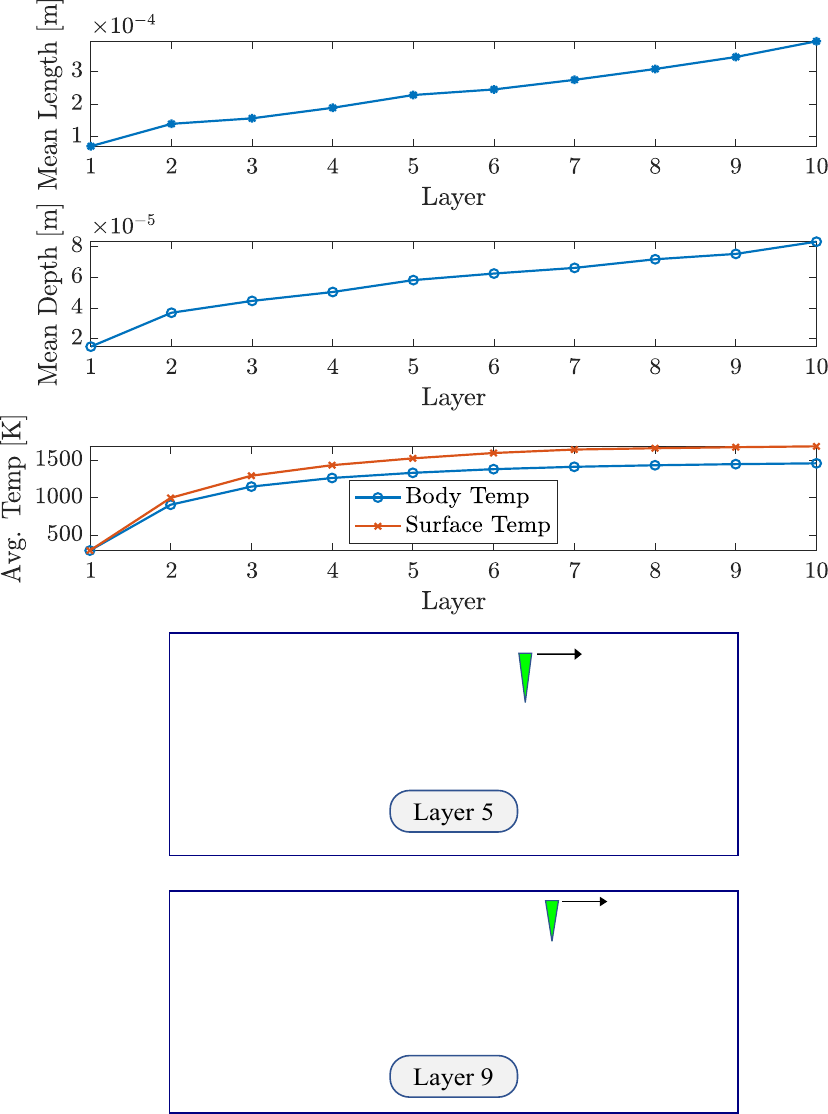}
    \caption{Print of a 10-layer part with constant 125W input. As each layer is printed the laser deposits thermal energy in the part. As this energy builds up, the melt pool becomes larger and more energetic leading to non-uniformities in the part. This heat buildup could be avoided by allowing the part to completely cool between layers but at the cost of significantly increasing the build time for the part.}
    \label{fig:heat-buildup}
\end{figure}

The melt pool characteristics govern the solidification behavior and cooling rates within the process, which in turn impact the microstructure and porosity of the printed component. For example, a particular problem when applying a constant laser power input is heat buildup over the course of a print. This additional energy in the part causes the melt pool to grow over time as depicted in Figure~\ref{fig:heat-buildup} which can lead to defects, cracking, and a non-uniform microstructure in the finished part.

\subsection{Control Architecture}

\begin{figure}[htbp]
	\centering
	\includegraphics[width=0.95\columnwidth]{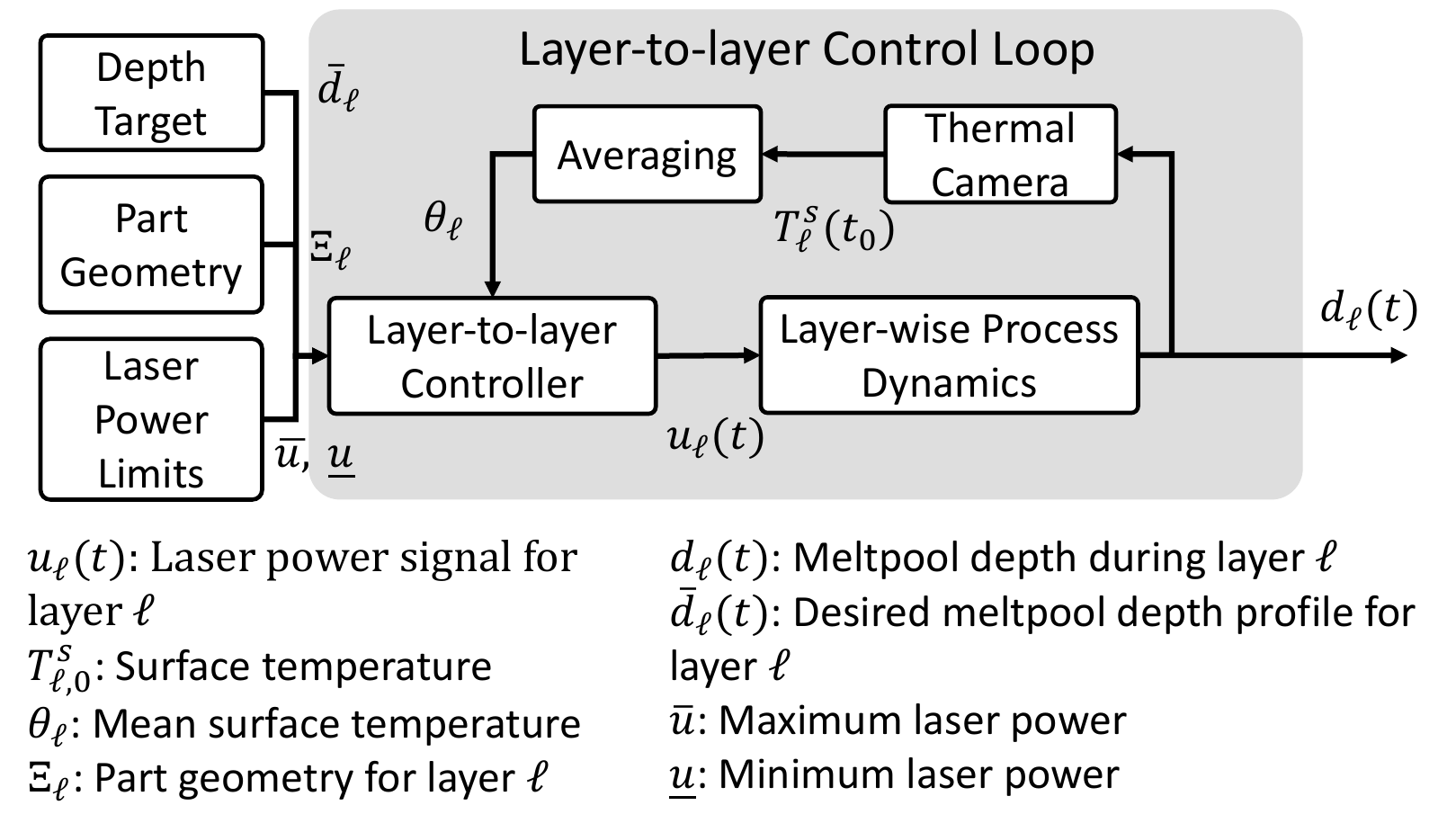}
	\caption{The proposed layer-to-layer control architecture}
\label{fig:l2l-arch}
\end{figure}

We propose an L2L control architecture for maintaining the melt pool depth at a desired set point by adjusting the laser power profile at each layer. Our detailed architecture is illustrated in Figure~\ref{fig:l2l-arch}. Before printing each layer we measure the surface temperature of the part and then compute a laser power profile for the layer by solving an optimal control problem.
The multi-layer process is indexed using $\ell \in \{0,\ldots,\mc{L}-1\}$, where $\mc{L}$ is the total number of layers in the part. In our setting, the laser moves at a constant speed $v$ along a 1D track with each layer taking a constant $\tau$ seconds. Our control input is the laser power $u_\ell(t)$ with minimum and maximum power levels $\underbar{u}$ and $\bar{u}$, respectively. We discretize the interval $[0,\tau]$ into $N$ evenly spaced intervals, leading to a discrete-time domain $\mathcal{T}= \{0,..., t^i, \ldots ,\tau \}$, and parameterize the laser power using a piecewise constant signal
\begin{equation}
	u_\ell(t) = \begin{cases}
		u^i_{\ell} & t\in [t^{i},t^{i+1}), \quad i \in \{0,\ldots,N-1\}\\
		u^{N-1}_\ell & t = \tau
	\end{cases}
\end{equation}
with powers $\mathbf{u}_\ell = [u^0_\ell,~\ldots~,u^{N-1}_\ell]^T$. The input signal in each layer is updated by the controller, which uses process measurements and a control-oriented model for optimization. We capture the geometry of the part using the sets $\Xi_\ell \subseteq \{0,\ldots,N\}$ which determine if the laser is on/off at different time points during each layer. By setting $u^i_\ell = 0$ if $i \notin \Xi_\ell$ and varying $\Xi_\ell$ over layers encodes layer-dependent geometries. 

\begin{figure}
    \centering
    \includegraphics[width=\columnwidth]{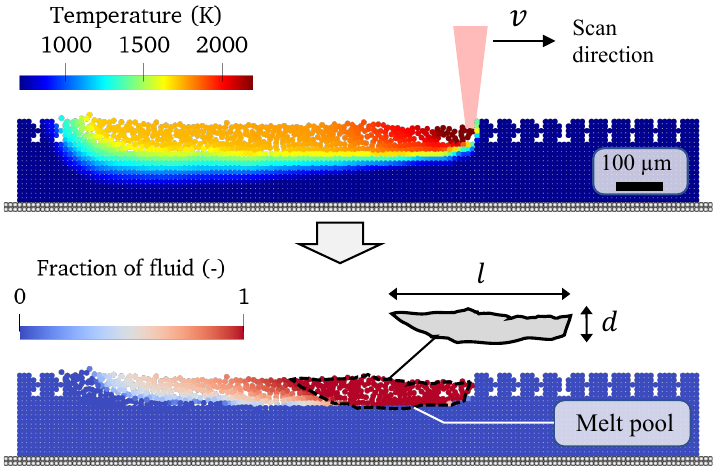}
    \caption{Snapshot of a high-fidelity LPBF simulation with the SPH method. Temperature distribution and fraction of fluid (FoF) are shown to visualize how the melt pool geometry can be extracted from the particles whose FoF is 1.}
    \label{fig:sim_melt pool}
\end{figure}

Our objective is to control the dimensions of the melt pool. The depth $d$ and length $l$ of the melt pool are extracted from the SPH simulation by drawing a bounding box around particles whose fraction of fluid is equal to one as illustrated in Figure~\ref{fig:sim_melt pool}.
We use the surface temperature of the part $T_{\ell}^s(0)$, measured at the beginning of each layer, as a feedback input. It serves as a proxy for the energy contained in the part and can be readily measured on real machines using thermal cameras (see~\cite{kavas2023layer}).

\begin{remark}
In this paper we focus on adjusting the laser power and consider the scan speed fixed. Another option to control the melt pool would be to adjust the cooling time, however this option is less flexible than adjusting the entire laser power profile since it would only offer control over the initial temperature of each layer.
\end{remark}

\subsection{Control-oriented Modeling}
Our controller computes the laser power input profile for each layer by solving an optimal control problem (OCP). Accurately controlling the melt pool dimensions requires a model that captures the response of the melt pool to changes in the laser power. Ideally, this model should capture the essential dynamic behaviour of the melt pool while remaining low-complexity enough to easily incorporate into a tractable optimal control problem.

For this purpose, we propose using a discrete-time reduced-order linear parameter varying (LPV) model of the form
\begin{equation}\label{eq:lpv_model}
    x^{i+1} = A(\theta)x^i + B(\theta)u^i + c(\theta)
\end{equation}
where $u^i$ [W] is discretized laser power, the state $x = [d~~l]^T$ consists of the melt pool depth $d$ [m] and length $l$ [m], $A(\theta), B(\theta)$ and $c(\theta)$ are the linear and affine terms respectively in our LPV model and the scheduling parameter $\theta$ is the average surface temperature of the preceding layer at the beginning of each layer as illustrated in Figure~\ref{fig:sim_melt pool}. 

The control-oriented model~\eqref{eq:lpv_model} predicts the evolution of the melt pool dimensions within a layer based on laser power input and information about the previous layers encapsulated in the average temperature parameter $\theta$. In Section~\ref{sec:sim} we demonstrate that the scheduling parameter $\theta$ effectively condenses previous layer information about the process. 

To obtain system identification data, we run the multi-layer simulation using randomized laser power input signals and obtain $\mathcal{K}$ collections of time-series data
\begin{equation}
    D = \{D_{k}\}_{k = 1}^{\mathcal{K}},~~ D_k = \{\{\hat x^i\}_{k}, \{\hat u^i\}_k, \theta_{k}\},
\end{equation}
where $D_k$ is simulated input-output data for a single layer. We then use this data to estimate the model parameters $A,B,c$ by solving the least-squares problem
\begin{equation}
    (A,B,c) = \argmin_{A,B,c} \sum_{i} \|\hat x^{i+1} - A \hat x^i - B\hat u^i - c\|^2
\end{equation}
which yields a collection of linear models $\{A(\theta_k),B(\theta_k),c(\theta_k)\}$ valid at the corresponding parameter values $\theta_k$. Whenever we need to evaluate $A,B$, and $c$ outside of our grid of parameter values $\theta_k$, we use linear interpolation. This yields a continuous function $A(\theta),B(\theta),c(\theta)$ that is valid within the domain $[\min_k \theta_k, \max_k \theta_k]$ that we can use to implement \eqref{eq:lpv_model}.

\subsection{Optimal Control Problem}
We compute the laser power profile for each layer by solving a trajectory optimization problem for each subsequent layer. Our high-level objective is to drive the melt pool depth to a target $\bar d$. The resulting optimal control problem is
\begin{subequations} \label{eq:ocp}
\begin{alignat}{3}
&\mc{P}(\bar d, \Xi, \theta) =&&\argmin_{u}\sum_{i = 0}^{N} \|d^{i} - \bar d\|_Q^2 + \|u^{i} - u^{i-1}\|_R^2 ~~\label{eq:ocp-cost}\\
&\quad \quad \mathrm{s.t.} &&\begin{bmatrix}
    x^{0} \\ u^{-1}
\end{bmatrix} = \begin{bmatrix}
    0\\ \underbar{u}
\end{bmatrix} \label{eq:ocp-ic} \\
& &&x^{i+1}= A(\theta) x^i + B(\theta) u^i + c(\theta), \label{eq:ocp-dynamics}\\
& && \nonumber \quad \quad \quad \quad \qquad \qquad \quad  \qquad i = \{0,\ldots,N-1\} \\
& &&u^{i} \in \begin{cases}
[\underbar u,\bar u] & \text{if~} t_i \in \Xi\\
\{0\}& \text{else}
\end{cases}, ~i = \{0,\ldots,N\} \label{eq:ocp-cstr}
\end{alignat}
\end{subequations}
where $i$ indexes time within the layer, $x^i = (d^i,l^i)$ are the melt pool dimensions, $\theta$ is the average surface temperature at the beginning of the layer, $\Xi$ encodes the layer geometry, $\underbar{u}$ and $\bar u$ are laser power bounds, and $Q,R \succeq 0$ are symmetric weighting matrices. The cost function \eqref{eq:ocp-cost} balances depth tracking with an incremental penalty that promotes a smooth power profile (desirable for the laser)\footnote{Various other control objectives could be employed in~\eqref{eq:ocp-cost} based on the application of interest, including melt pool length, temperature, or combinations of all variables. Here we focus on melt pool depth as it has a profound impact on the resulting print quality, and leave exploring other control objectives for future work.}. The initial conditions \eqref{eq:ocp-ic} encodes that there is no initial melt pool and gives an initial value for the incremental penalty while the dynamics \eqref{eq:ocp-dynamics} model how the melt pool dimensions respond to the laser power input. The model \eqref{eq:ocp-dynamics}  depends on the average initial surface temperature $p$ which is how data from the process enters the OCP. Finally the input constraints \eqref{eq:ocp-cstr} encode the geometry of the part and enforce physical limits on the laser power.

\begin{figure}[htbp]
    \centering
    \includegraphics[width=0.95\columnwidth]{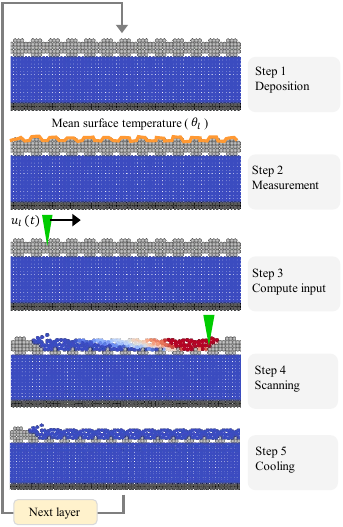}
    \caption{Integration of the L2L controller into the multi-layer simulation process. Lines 2, 3, and 4-6 of Algorithm~\ref{algo:l2l} correspond to steps 2, 3, and 4, respectively.}
    \label{fig:tasks}
\end{figure}

The overall closed-loop control algorithm employed in this work is summarized in Algorithm~\ref{algo:l2l} and its integration into the high-fidelity simulator is illustrated in Figure~\ref{fig:tasks}. The next section provides further implementation details specific to the process parameters and geometries we considered in this work. 
\begin{algorithm}[h]
	\caption{Layer-to-Layer Feedback}
	\label{algo:l2l}
	\begin{algorithmic}[1]
		\renewcommand{\algorithmicrequire}{\textbf{Inputs:}}
		\renewcommand{\algorithmicensure}{\textbf{Outputs:}}
		\Require Target depth $\bar d$, layer geometries $\Xi_{\ell}$, bounds $[\underbar{u},\bar u]$
		\For{Each layer $\ell \in [0,\ldots,\mc{L}]$}
		\State Measure the surface temperature field $T^{0,s}_{\ell}$ and compute its (spatial) average $\theta$
		\State Solve the OCP \eqref{eq:ocp} to determine the power input profile $ \{u^i_*\}_{i = 0}^{N-1} = \mc{P}(\bar d,\Xi_\ell,\theta)$
		\For{each timestep $t^i \in \{0,\ldots,t^{N-1}\}$ in layer $\ell$}
		\State Set the laser power $u(t) = u^i_*$ for $t\in [t^{i},t^{i+1})$
		\EndFor
		\EndFor
	\end{algorithmic}
\end{algorithm}

\section{Numerical Experiments}
\label{sec:sim}
In this section, we illustrate the performance of the proposed closed-loop control method with high-fidelity simulations to control the depth of an LPBF process. We first outline the simulation setup, geometries used in the study, and simulation-specific parameters before giving details about control model identification. Finally, closed-loop simulation results for two geometries are provided and we study the temperature and effect of different geometries by drawing parallels to experimental results from real processes.

\subsection{Simulation Setup}
\label{sec:sim_setup}
We consider a test case of scanning a 2-mm long and 0.15-mm high substrate made of 304L stainless steel with the material properties given in \cite{russell2018numerical}. Powder particles of the same material with a diameter of 40 $\mu$m sit uniformly on the substrate at the first layer. The laser beam has a 40 $\mu$m radius at its focal point and scans the track at a constant speed of 1500 mm/s. The cooling time between layers is set as 1.5x the scanning time. A laser absorption coefficient of 0.27 as in \cite{russell2018numerical,afrasiabi2021multi}, is used in the simulations. We set $\underbar u = 100$ W which is the minimum laser power needed to consistently obtain a melt pool and $\bar u = 150$ W as the maximum laser power. We set $N = 40$ throughout.

The high-fidelity model of this LPBF process was created by discretizing the substrate and powder grains with SPH particles of uniform size, we chose a modest spatial resolution of approximately $\Delta x=7\;\mu$m as the initial particle spacing with a smoothing length of $h=1.3\Delta x$. Validation results for our 2D SPH-based model of a single-layer LPBF process against the experimental melt pool dimension measurements of \cite{khairallah2014mesoscopic} can be found in \cite{afrasiabi2021multi} for the same material.

\begin{figure}
    \centering
    \includegraphics[width=0.9\columnwidth]{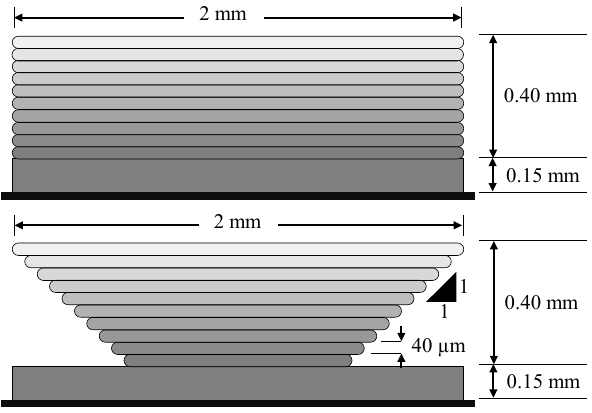}
    \caption{The brick (left) and overhang (right) geometries considered in the closed-loop simulation studies.}
    \label{fig:geom_def}
\end{figure}

Figure \ref{fig:geom_def} shows the two geometries considered in the closed-loop simulations. We select a simple block geometry as a baseline and more challenging overhanging trapezoid. We use the trapezoid geometry to validate the generalization capabilities of our control architecture with an LPV model identified on the simple block geometry. Overhang geometries are of particular interest since they may require additional support structures that may be costly to build in practice. Furthermore, due to the increasing thermal mass in subsequent layers, overhang geometries are prone to overheating~\cite{kavas2023layer} since heat cannot easily flow into the base plate\footnote{Conductive heat transfer through the solidified part into the base plate is the dominant cooling mechanism in LPBF processes.}.

The L2L controller is implemented in \texttt{C++} and linked directly with the SPH simulation code. The OCP \eqref{eq:ocp} forms the core of the controller and is a quadratic program, a class of optimization problems that can be solved efficiently and reliably. We solve it using a \texttt{C++} implementation\footnote{\url{https://github.com/dliaomcp/fbstab}} of the FBstab algorithm \cite{liao2020fbstab}. 

\subsection{Controller Model Identification} \label{ss:generation-data}
\begin{figure}[h]
	\centering
	\includegraphics[width=\columnwidth]{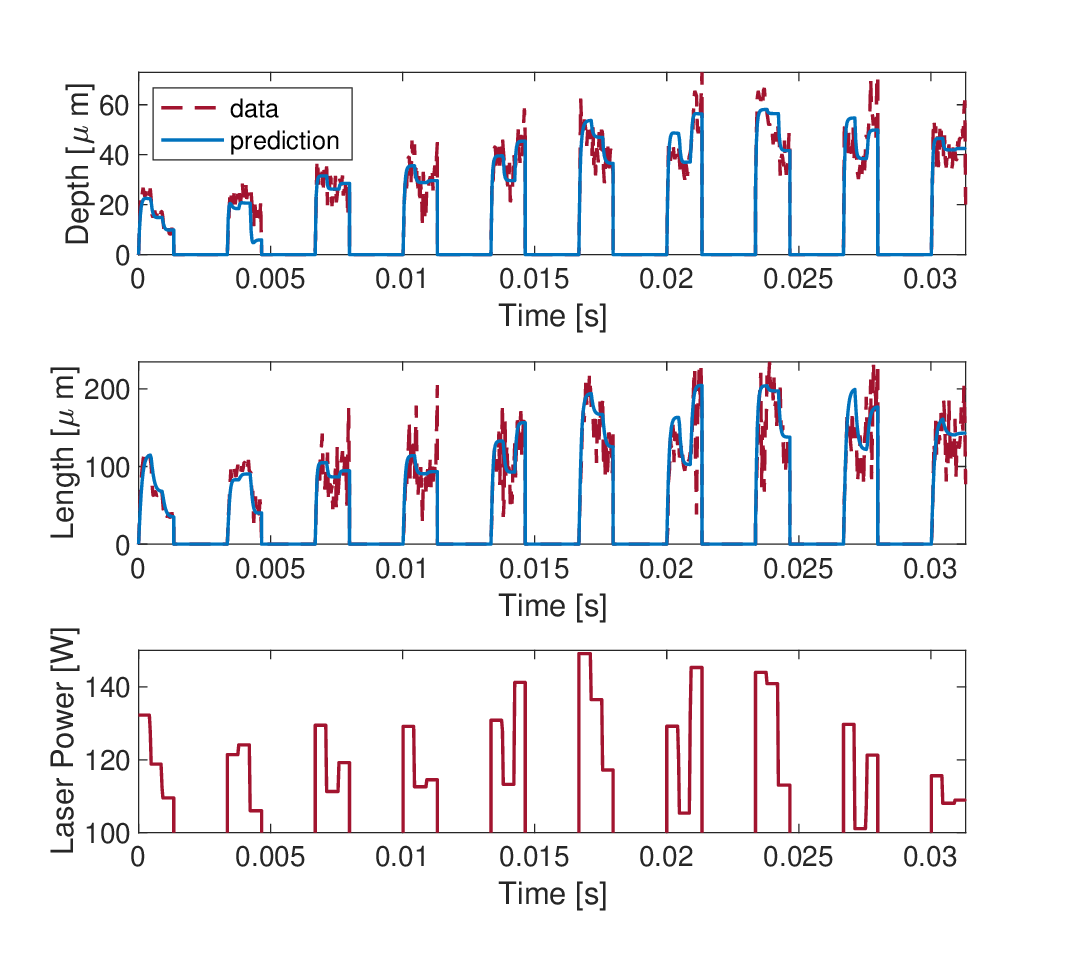}
	\caption{Our proposed control-orientated LPV model captures the low-frequency dynamics of the melt pool well. The melt pool dimensions become larger and noisier over time and energy accumulates in the part. Each of the plot sections indicates a layer.}
	\label{fig:model-validation}
\end{figure}

To train the model~\eqref{eq:lpv_model} we generate system identification data by applying random piecewise constant laser input signals to an $\mc{L} = 10$ layer print with a ``block'' geometry. Specifically, we divide each layer into three equal sections and apply a piecewise constant identification signal $u(t) = \{u_j~~t\in (t_j,t_{j+1}],~~j \in \{0,1,2\}\}$ where the laser power levels are sampled from the uniform distribution $u_j \sim \mathcal{U}[100,150]$. We use only three intervals per layer to limit excitation of high-frequency dynamics. Examples of the generated laser power input profile are displayed in the third row of Figure~\ref{fig:model-validation}. We simulate twenty ten-layer prints (i.e., $n_e = 20$ $\mc{L} = 10$) to form our training data set. We observed mean surface temperatures in the range $[293~~1340]$ K. 

The melt pool dimensions predicted by the model are compared using validation data (obtained from independent simulations generated in the same way as the training dataset with a mean temperature range $[293~~1280]$ K) in Figure~\ref{fig:model-validation}. The model captures the essential low-frequency components of the dynamics well. The melt pool increases in size and becomes noisier as the layers progress and as energy accumulates in the part. The root mean square length and prediction errors for both training and validation datasets are presented in Table~\ref{tab:rsme}, the error for the validation data is similar to that of the training data, indicating good generalization performance. These results indicate that a relatively simple scheduling parameter that uses the directly measurable surface temperature adequately captures the low-frequency system dynamics of the melt pool in the 2D case. 

\begin{table}[t]
	\centering
	\caption{The root mean square error (RSME) between the data and prediction of the model are similar for the training and validation datasets} \label{tab:rsme}
	 \renewcommand{\arraystretch}{1.25}   
	\begin{tabular}{@{}c c c @{}} 
		\toprule
		& Training data & Validation data \\ 
		\cmidrule(lr){2-3}
		Depth RSME [$\mu$m]  & 3.71 & 3.38 \\ 
		Length RSME [$\mu$m] & 15.81 & 15.87 \\
		\bottomrule
	\end{tabular}
\end{table}

\begin{figure*}
	\centering
	\includegraphics[width=1.6\columnwidth]{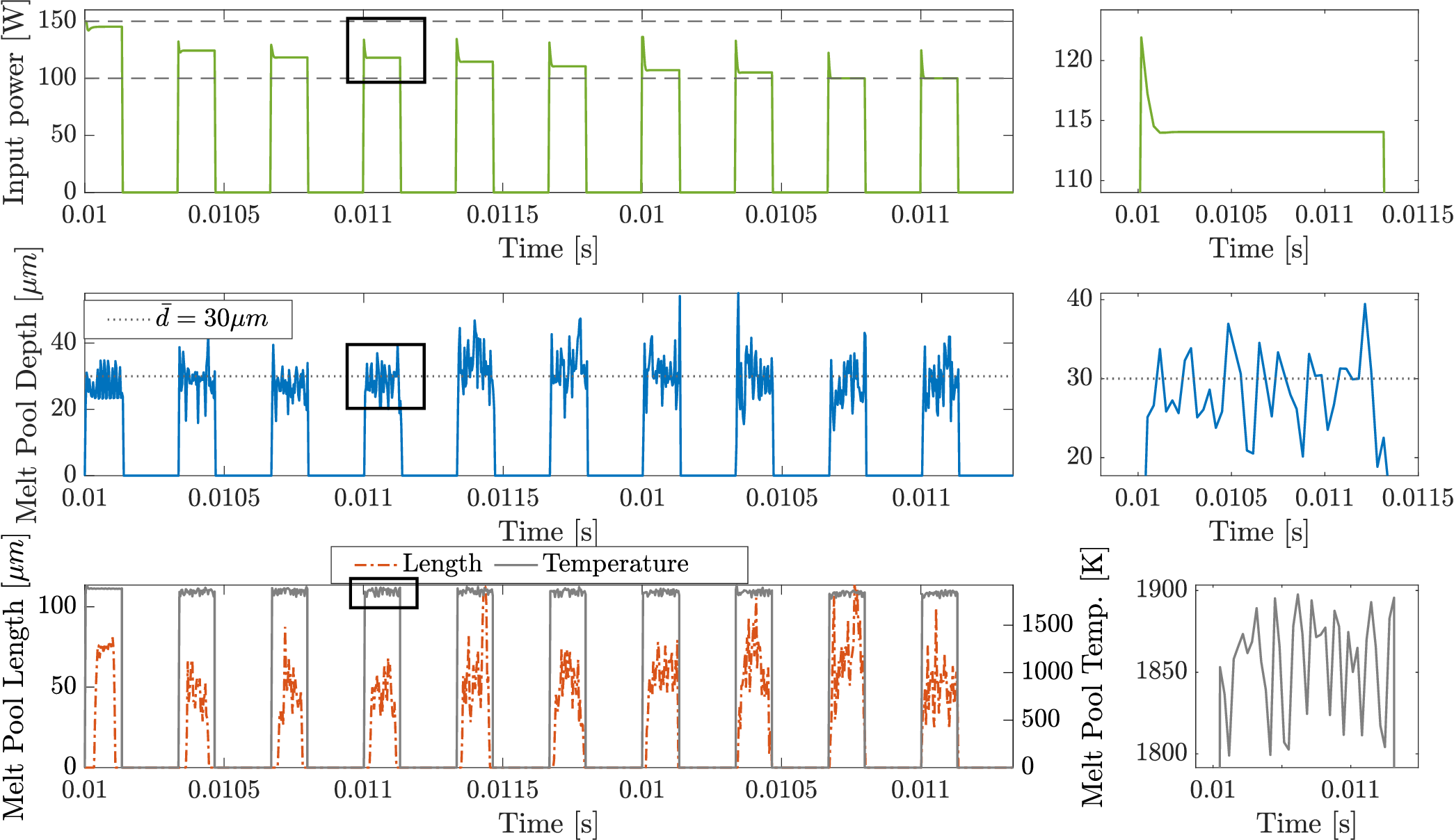}
	\caption{Closed-loop control results for the brick geometry. Top figure is the input signal for the print with the dashed horizontal lines denoting input limits. Middle plot shows the melt pool depth with the reference value as a dotted line. Bottom plot shows the melt pool length on the left vertical axis and the melt pool temperature on the right vertical axis. The gaps in the data between layers show the recoating time. The figures on the right column are zoomed segments of those on the left column marked with black rectangles.}
	\label{fig:brick_results}
\end{figure*}

\begin{figure*}
	\centering
	\includegraphics[width=1.6\columnwidth]{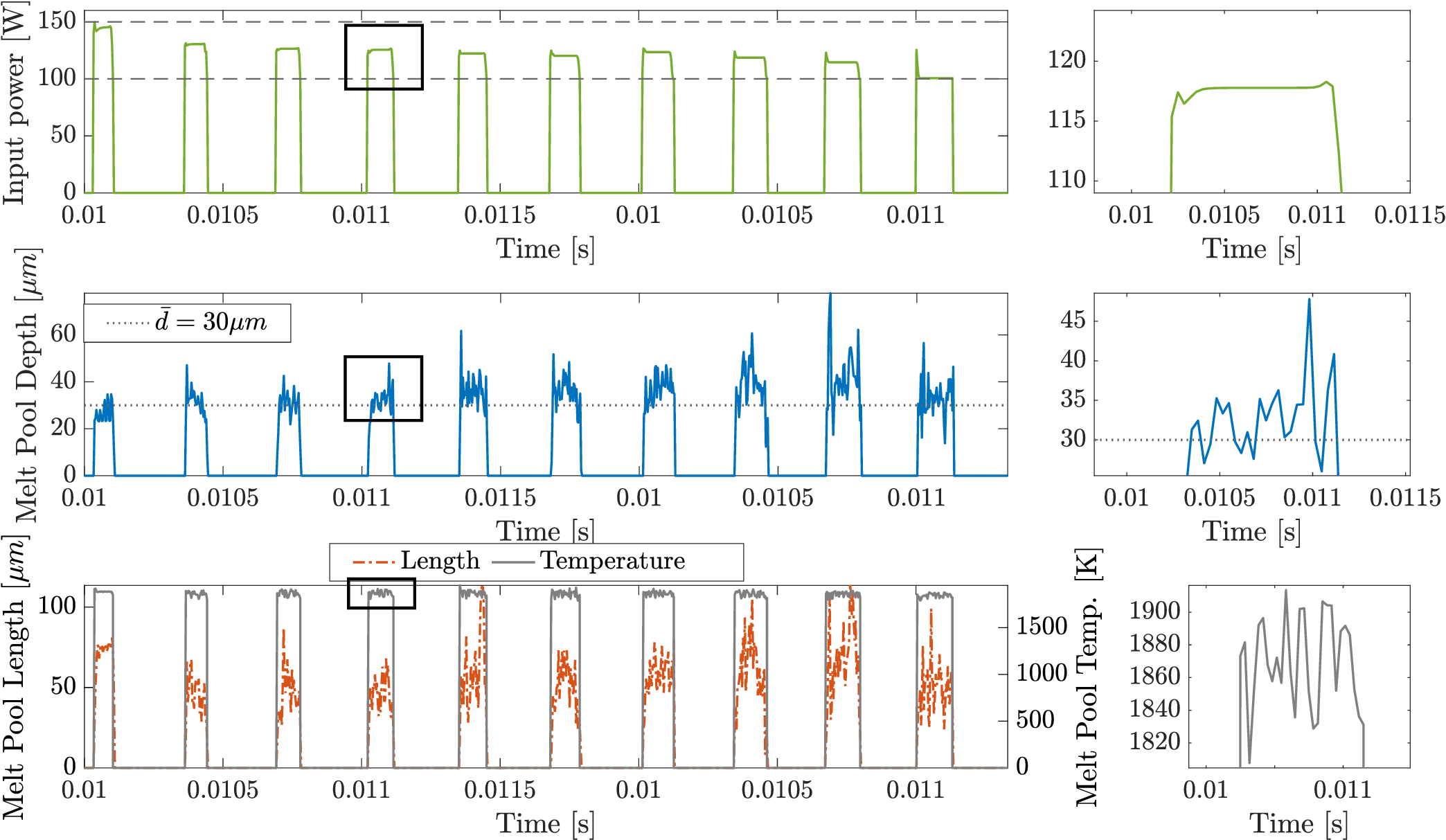}
	\caption{Closed-loop control results for the overhang geometry. Top figure is the input signal for the print with the dashed horizontal lines denoting input limits. Middle plot shows the melt pool depth with the reference value as a dotted line. Bottom plot shows the melt pool length on the left vertical axis and the melt pool temperature on the right vertical axis. The gaps in the data between layers show the recoating time. The figures on the right column are zoomed segments of those on the left column marked with black rectangles.}
	\label{fig:overhang_results}
\end{figure*}

\begin{figure*}[h]
	\centering
	\begin{subfigure}{0.49\textwidth}
		\includegraphics[width=\textwidth]{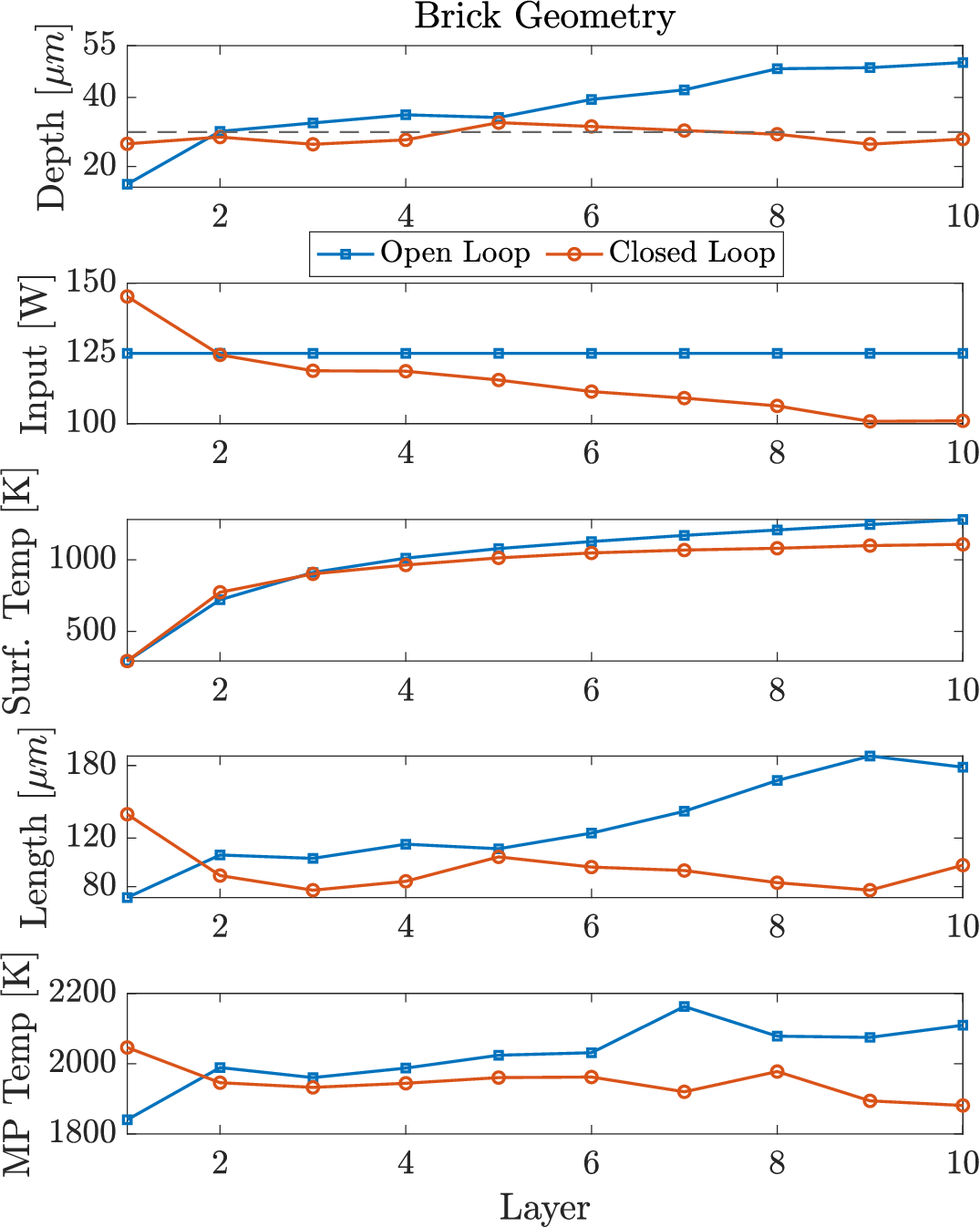}
		\label{fig:brick_average_results}
	\end{subfigure}
	\hfill
	\begin{subfigure}{0.49\textwidth}
		\includegraphics[width=\textwidth]{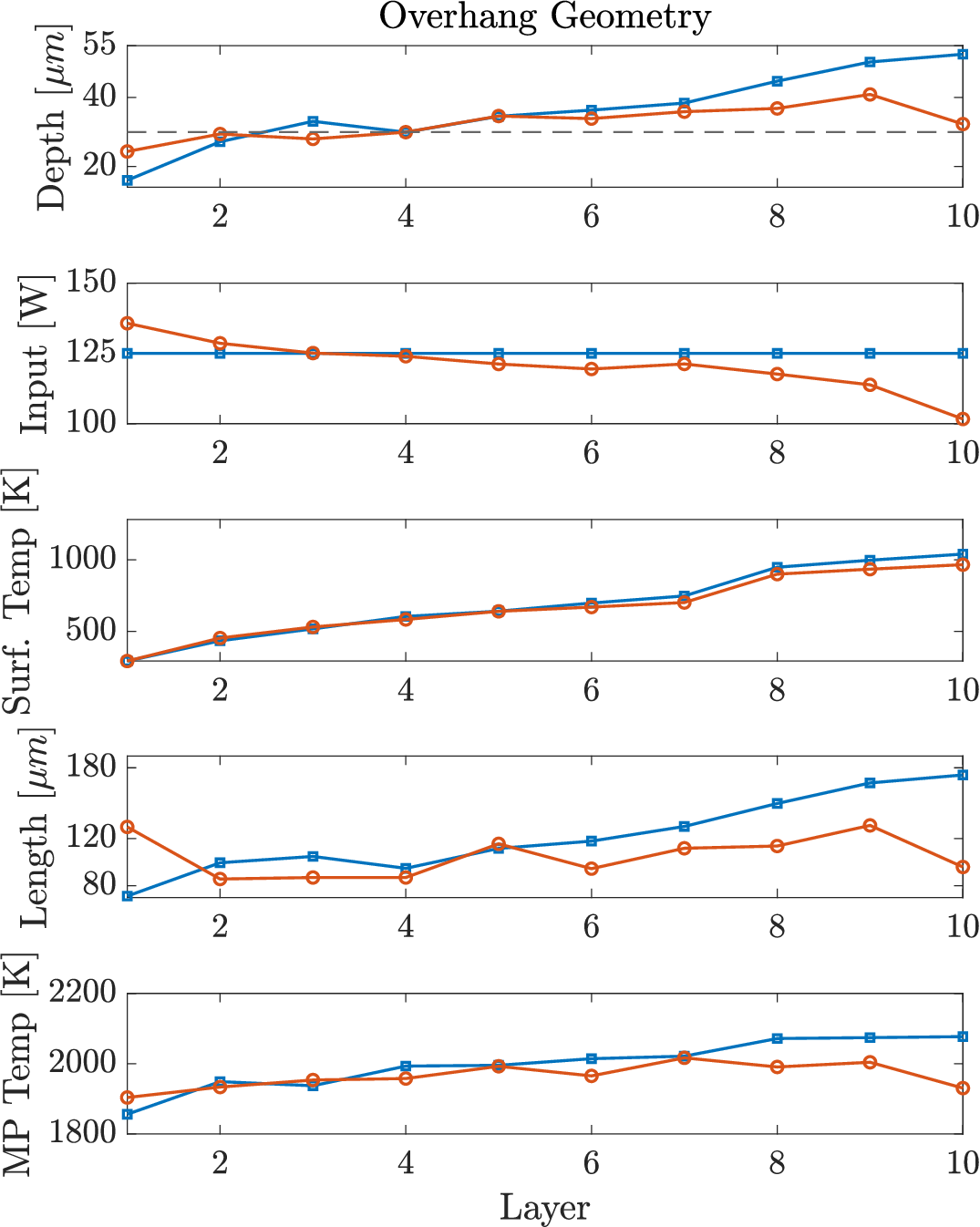}
		\label{fig:overhang_average_results}
	\end{subfigure}
	\caption{Closed- and open-loop control results for the brick (left) and overhang (right) geometries. Data points show the average for a given layer. Surface temperature data is the average temperature of the layer surface at the end of the laser pass. The dashed horizontal line shows the target depth of $30~\mu m$.}
	\label{fig:olcl_combined}
\end{figure*}

\subsection{Depth Tracking Results with L2L Control}
\label{sec:res}
Figures~\ref{fig:brick_results} - \ref{fig:overhang_results} illustrate the performance of our controller in closed-loop with the SPH model. In both cases, the L2L controller with the identified LPV model tracks the reference depth at each layer while satisfying the desired input constraints.
To compensate for a buildup of thermal energy in the part, the controller reduces the input at each subsequent layer to stabilize the melt pool depth at the desired target of $30\mu m$. Within each layer, the controller initially applies a high input for a short amount of time to get the depth near the desired reference, then reduces the input and keeps it roughly constant for the rest of the layer. In both figures, the plots in the bottom row show the behavior of the melt pool length and temperature. These properties are not explicitly tracked by the controller as the only objective is to track the reference depth. However, due to the physics of the process, tracking a reference depth indirectly also stabilizes the melt pool length and temperature. 

Recall that the model was identified using data from the brick geometry. Surprisingly, when the same model is used to control the overhang geometry the results are qualitatively similar to the brick geometry despite the model mismatch, suggesting that our controller can successfully handle at least a limited range of geometries. Quantitative differences do of course exist; for example, the shape of the input signals is more variable and the layer-average power input does not decrease monotonically as in the block case. This shows the L2L controller adapting to the new geometry through feedback. 

\subsection{Comparison between open and closed-loop}

Figure~\ref{fig:olcl_combined} compares the closed-loop results with the performance of an open-loop strategy that applies a constant 125 W signal to each layer. The open-loop strategy increases the temperature and melt pool dimensions from one layer to the next as energy builds up in the part. The closed-loop controller on the other hand avoids this by gradually decreasing the average input power at each layer. The average depth graphs demonstrate again the impact of model mismatch on controller performance, In the overhang geometry, the controller keeps the depth close to the desired level and decreases the input aggressively between layers 9 and 10, which can be explained by the excessive error at layer 9 (also reflected on the surface temperature as previously shown).

\begin{figure*}[h]
	\centering
	\includegraphics[width=\textwidth]{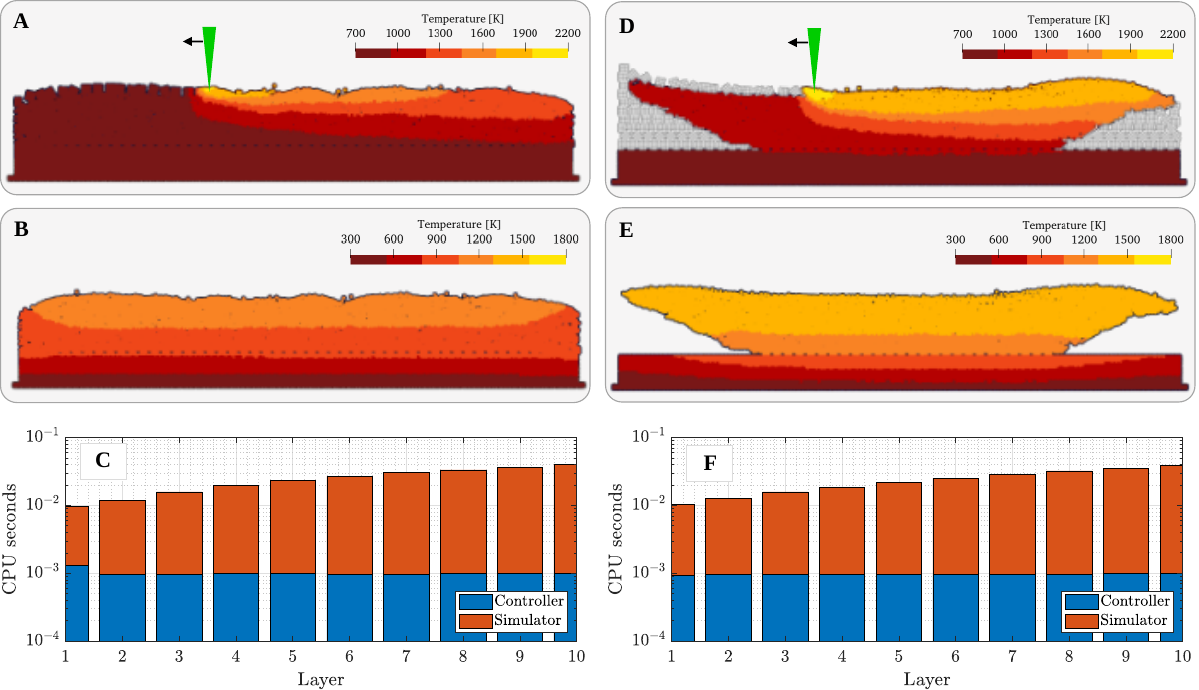}
	\caption{\textbf{Top and Middle Rows} Temperatures of the LPBF-ed part at layer 10 predicted by the SPH simulation: A) Brick after 30.9 ms, when the laser is scanning the powder bed; B) Brick after 33.3 ms, when the laser is off and the process is finished, D) Overhang after 30.9 ms when the laser is scanning the powder bed; E) Overhang after 33.3 ms when the laser is off and the process is finished
		\textbf{Bottom Row} - Runtime of one full calculation loop for each module per layer in different LPBF geometries: C) Block rectangular; F) Overhang pyramid. The "simulator" gets more expensive layer by layer while the "controller" remains indifferent.}
	\label{fig:sim_comb}
\end{figure*}

\subsection{Runtime}
The full 10-layer LPBF simulation and integrated L2L controller module described in Section \ref{sec:sim_setup} takes about 2 hours on a single core of AMD Ryzen™ 7 3700X at 3.60 GHz. Figure~\ref{fig:sim_comb}C and Fig.~\ref{fig:sim_comb}F compare the closed-loop simulation times for the two geometries. The simulation times do not vary appreciably by geometry. The reported CPU seconds are averaged over the activation period to indicate a single representative value for each module. The L2L controller executes about one order of magnitude faster than the simulator; moreover, unlike the simulator, its runtime is independent of the layer as the model only captures the top layer. The controller run-time hovers around a millisecond which is roughly an order of magnitude smaller than the layer printing time, suggesting it could be implemented in real-time. It is slightly larger during the first layer due to constructors and remains roughly constant thereafter. In this paper all experiments use $N = 40$ but the algorithm can be readily scaled. The runtime of the QP solver we employ scales linearly with $N$ thanks to tailored matrix factorization routines based on Riccati recursions as discussed in \cite[Section 5]{liao2020fbstab}.

\subsection{Geometry and Temperature}
A salient advantage of the high-fidelity simulator is access to all the physical variables of the process. Figure \ref{fig:sim_comb} compares the geometry and temperature distribution of the final parts. The heat-affected zones in Figure \ref{fig:sim_comb}A and Figure \ref{fig:sim_comb}D evolve as the laser beam travels from right to left, demonstrating the approximate location of the melt pool region. In the overhang geometry case, more fluid motion follows the scanning laser trail that, in turn, leads to non-symmetric and skewed down-facing areas at both ends. Furthermore, the final temperature fields (Figure \ref{fig:sim_comb}B and Figure \ref{fig:sim_comb}E) exhibit a layer-wise distribution pattern, implying that the choice of cooling time between layers was sufficient to reach a uniform surface temperature. We hypothesize the near uniformity of the surface temperature is why the predictive LPV model \eqref{eq:lpv_model} used in the controller is able to effectively capture the impact of previous layers using a single average temperature parameter.

When the LPBF process is complete, the final layer's surface of the block geometry is colder than the overhang pyramid - see the yellowish areas in Figure \ref{fig:sim_comb}B vs. Figure \ref{fig:sim_comb}E. The cause of this behavior is additional mass at the lower layers of the brick geometry that increases heat diffusion, as shown experimentally in \cite{kavas2023layer}.

\section{Outlook for implementation on a real system}
\label{sec:outlook}

\subsubsection{Sensors and Data} Our controller architecture is designed with existing sensor technology in mind. The controller requires the average surface temperature of the part at the beginning of each layer which can be readily obtained from a thermal camera, a relatively inexpensive piece of equipment that can be implemented in the build chamber. For example, such a camera was used in \cite{kavas2023layer} for surface temperature stabilization. 

The data needed to train the model can be obtained by CMOS camera measurements of the melt pool shape or pyrometer data to estimate the length of the melt pool in real-time. Obtaining the depth is more difficult from in-situ data, but can be estimated through destructive testing of the finished parts and single-track experiments~\cite{yadroitsev2010single}. Additionally, in-situ measurements can be utilized to estimate melt pool depth with high accuracy through empirical correlations~\cite{trapp2017situ}.

\subsubsection{Real-time execution}
The layer-to-layer update and recoating times in real systems are significantly larger than those reported in our study, therefore we expect our method to be feasible for real-time implementation.

\subsubsection{Integration with in-layer controllers}
Our layer-to-layer directly provides a time-domain reference signal and can thus be readily integrated with within-layer controllers such as \cite{benda1994temperature,renken2019process,Renken2018,shkoruta2022real,wang2020model} by using the melt pool length or temperature computed by the L2L controller as a reference signal. This hierarchical implementation may improve the efficiency of the closed-loop process since the low-level within-layer controllers stabilize the melt pool dimensions about a given reference. Such applications are sought after in practice and are an important direction for future work.

\subsubsection{Extension to 3D processes}

In this paper, we use an LPV surrogate model for the process dynamics that exploits that the surface temperature typically is near uniform when printing 2D geometries. However in 3D, thermal dynamics between adjacent scan vectors and common hatching strategies that change the direction in each layer could degrade the accuracy of this surrogate. Nevertheless, the proposed method in this paper is expected to perform well on print conditions with longer scan vectors or identical vector scan directions in a 3D geometry, such that vector-to-vector interactions are limited.

Furthermore, the proposed control methodology can incorporate a variety of process models and could be extended to 3D by identifying an appropriate (and likely more complex) surrogate model for the process dynamics. E.g., using a neural network and including the entire surface temperature field from the previous layer or employing a spatial representation of the input dynamics instead of a temporal one could help incorporate hatch orientation as different print paths\footnote{In the worst case of a solid hatch, the computational complexity of our trajectory-based discretization would grow with $N^2$ but would be much smaller for sparser patterns.}, similar to~\cite{liao2022layer}.

\section{Conclusion}
\label{sec:conclusion}
This paper proposes a layer-to-layer (L2L) controller for melt pool control in LPBF and illustrates its performance in closed-loop with a high-fidelity simulation of the process. The controller is able to successfully control the melt pool and is shown to be robust with respect to changing geometries. Our approach provides insights and analysis tools that would not be possible with experimental setups.

This study provides the first steps in analyzing L2L controllers with high-fidelity simulations to understand the efficacy of different control structures for melt pool control, which opens up new research questions for future studies. Some of the immediate future directions include a study of combined within-layer and L2L control architectures to understand the performance trade-offs, further studies on different control architectures such as multi-layer model predictive control, the impact of varying geometries, studies using different measurement inputs in conjunction with appropriate observers for the controller, and the investigation of more complex surrogate models suitable for 3D processes.

\bibliography{slm}           

\begin{thebibliography}{10}

\bibitem{dowling2020review}
L.~Dowling, J.~Kennedy, S.~O'Shaughnessy, and D.~Trimble, ``A review of
  critical repeatability and reproducibility issues in powder bed fusion,''
  {\em Materials \& Design}, vol.~186, p.~108346, 2020.

\bibitem{ahmed2022process}
N.~Ahmed, I.~Barsoum, G.~Haidemenopoulos, and R.~A. Al-Rub, ``Process parameter
  selection and optimization of laser powder bed fusion for 316l stainless
  steel: A review,'' {\em Journal of Manufacturing Processes}, vol.~75,
  pp.~415--434, 2022.

\bibitem{mani2017review}
M.~Mani, B.~M. Lane, M.~A. Donmez, S.~C. Feng, and S.~P. Moylan, ``A review on
  measurement science needs for real-time control of additive manufacturing
  metal powder bed fusion processes,'' {\em International Journal of Production
  Research}, vol.~55, no.~5, pp.~1400--1418, 2017.

\bibitem{tapia2014review}
G.~Tapia and A.~Elwany, ``A review on process monitoring and control in
  metal-based additive manufacturing,'' {\em Journal of Manufacturing Science
  and Engineering}, vol.~136, no.~6, 2014.

\bibitem{benda1994temperature}
J.~Benda, ``Temperature-controlled selective laser sintering,'' in {\em 1994
  International Solid Freeform Fabrication Symposium}, 1994.

\bibitem{kruth2007feedback}
J.-P. Kruth, P.~Mercelis, J.~Van~Vaerenbergh, and T.~Craeghs, ``Feedback
  control of selective laser melting,'' in {\em Virtual and Rapid
  Manufacturing}, pp.~521--528, Crc Press, 2007.

\bibitem{shkoruta2022real}
A.~Shkoruta, S.~Mishra, and S.~J. Rock, ``Real-time image-based feedback
  control of laser powder bed fusion,'' {\em ASME Letters in Dynamic Systems
  and Control}, vol.~2, no.~2, 2022.

\bibitem{kavas2023layer}
B.~Kavas, E.~C. Balta, M.~Tucker, A.~Rupenyan, J.~Lygeros, and M.~Bambach,
  ``Layer-to-layer closed-loop feedback control application for inter-layer
  temperature stabilization in laser powder bed fusion,'' {\em Additive
  Manufacturing}, p.~103847, 2023.

\bibitem{heralic2012height}
A.~Herali{\'c}, A.-K. Christiansson, and B.~Lennartson, ``Height control of
  laser metal-wire deposition based on iterative learning control and 3d
  scanning,'' {\em Optics and lasers in engineering}, vol.~50, no.~9,
  pp.~1230--1241, 2012.

\bibitem{hagqvist2015resistance}
P.~Hagqvist, A.~Herali{\'c}, A.-K. Christiansson, and B.~Lennartson,
  ``Resistance based iterative learning control of additive manufacturing with
  wire,'' {\em Mechatronics}, vol.~31, pp.~116--123, 2015.

\bibitem{xia2020review}
C.~Xia, Z.~Pan, J.~Polden, H.~Li, Y.~Xu, S.~Chen, and Y.~Zhang, ``A review on
  wire arc additive manufacturing: Monitoring, control and a framework of
  automated system,'' {\em Journal of manufacturing systems}, vol.~57,
  pp.~31--45, 2020.

\bibitem{mu2022layer}
H.~Mu, J.~Polden, Y.~Li, F.~He, C.~Xia, and Z.~Pan, ``Layer-by-layer
  model-based adaptive control for wire arc additive manufacturing of thin-wall
  structures,'' {\em Journal of Intelligent Manufacturing}, vol.~33, no.~4,
  pp.~1165--1180, 2022.

\bibitem{sammons2013height}
P.~M. Sammons, D.~A. Bristow, and R.~G. Landers, ``Height dependent laser metal
  deposition process modeling,'' {\em Journal of Manufacturing Science and
  engineering}, vol.~135, no.~5, p.~054501, 2013.

\bibitem{sammons2018repetitive}
P.~M. Sammons, M.~L. Gegel, D.~A. Bristow, and R.~G. Landers, ``Repetitive
  process control of additive manufacturing with application to laser metal
  deposition,'' {\em IEEE Transactions on Control Systems Technology}, vol.~27,
  no.~2, pp.~566--575, 2018.

\bibitem{debroy2018additive}
T.~DebRoy, H.~Wei, J.~Zuback, T.~Mukherjee, J.~Elmer, J.~Milewski, A.~M. Beese,
  A.~d. Wilson-Heid, A.~De, and W.~Zhang, ``Additive manufacturing of metallic
  components--process, structure and properties,'' {\em Progress in Materials
  Science}, vol.~92, pp.~112--224, 2018.

\bibitem{bajaj2020steels}
P.~Bajaj, A.~Hariharan, A.~Kini, P.~K{\"u}rnsteiner, D.~Raabe, and E.~A.
  J{\"a}gle, ``Steels in additive manufacturing: A review of their
  microstructure and properties,'' {\em Materials Science and Engineering: A},
  vol.~772, p.~138633, 2020.

\bibitem{renken2019process}
V.~Renken, A.~von Freyberg, K.~Sch{\"u}nemann, F.~Pastors, and A.~Fischer,
  ``In-process closed-loop control for stabilising the melt pool temperature in
  selective laser melting,'' {\em Progress in Additive Manufacturing}, vol.~4,
  pp.~411--421, 2019.

\bibitem{Renken2018}
V.~Renken, L.~L{\"{u}}bbert, H.~Blom, A.~{Von Freyberg}, and A.~Fischer,
  ``{Model assisted closed-loop control strategy for selective laser
  melting},'' {\em Procedia CIRP}, vol.~74, pp.~659--663, 2018.

\bibitem{wang2020model}
Q.~Wang, P.~P. Michaleris, A.~R. Nassar, J.~E. Irwin, Y.~Ren, and C.~B.
  Stutzman, ``Model-based feedforward control of laser powder bed fusion
  additive manufacturing,'' {\em Additive Manufacturing}, vol.~31, p.~100985,
  2020.

\bibitem{asadi2021gaussian}
F.~Asadi, A.~Olleak, J.~Yi, and Y.~Guo, ``Gaussian process (gp)-based learning
  control of selective laser melting process,'' in {\em 2021 American Control
  Conference (ACC)}, pp.~508--513, IEEE, 2021.

\bibitem{wang2020layer}
X.~Wang, C.~S. Lough, D.~A. Bristow, R.~G. Landers, and E.~C. Kinzel, ``A
  layer-to-layer control-oriented model for selective laser melting,'' in {\em
  2020 American Control Conference (ACC)}, pp.~481--486, IEEE, 2020.

\bibitem{spector2018passivity}
M.~J. Spector, Y.~Guo, S.~Roy, M.~O. Bloomfield, A.~Maniatty, and S.~Mishra,
  ``Passivity-based iterative learning control design for selective laser
  melting,'' in {\em 2018 Annual American Control Conference (ACC)},
  pp.~5618--5625, IEEE, 2018.

\bibitem{zuliani2022batch}
R.~Zuliani, E.~C. Balta, A.~Rupenyan, and J.~Lygeros, ``Batch model predictive
  control for selective laser melting,'' in {\em 2022 European Control
  Conference (ECC)}, pp.~1560--1565, IEEE, 2022.

\bibitem{liao2022layer}
D.~Liao-McPherson, E.~C. Balta, R.~W{\"u}est, A.~Rupenyan, and J.~Lygeros,
  ``In-layer thermal control of a multi-layer selective laser melting
  process,'' in {\em 2022 European Control Conference (ECC)}, pp.~1678--1683,
  IEEE, 2022.

\bibitem{inyang2022model}
U.~Inyang-Udoh, R.~Hu, S.~Mishra, J.~Wen, and A.~Maniatty, ``Model-free
  multi-objective iterative learning control for selective laser melting,'' in
  {\em 2022 American Control Conference (ACC)}, pp.~2879--2885, IEEE, 2022.

\bibitem{rezaeifar2021line}
H.~Rezaeifar and M.~Elbestawi, ``On-line melt pool temperature control in l-pbf
  additive manufacturing,'' {\em The International Journal of Advanced
  Manufacturing Technology}, vol.~112, pp.~2789--2804, 2021.

\bibitem{vasileska2022novel}
E.~Vasileska, A.~G. Demir, B.~M. Colosimo, and B.~Previtali, ``A novel paradigm
  for feedback control in lpbf: Layer-wise correction for overhang
  structures,'' {\em Advances in Manufacturing}, vol.~10, no.~2, pp.~326--344,
  2022.

\bibitem{park2022melt}
B.~Park, X.~Wang, S.~Mishra, D.~Bristow, and R.~Landers, ``Melt pool feature
  extraction, spatial registration, and layer-to-layer control in powder bed
  fusion,'' in {\em 2022 International Solid Freeform Fabrication Symposium},
  2022.

\bibitem{king2015laser}
W.~E. King, A.~T. Anderson, R.~M. Ferencz, N.~E. Hodge, C.~Kamath, S.~A.
  Khairallah, and A.~M. Rubenchik, ``Laser powder bed fusion additive
  manufacturing of metals; physics, computational, and materials challenges,''
  {\em Applied Physics Reviews}, vol.~2, no.~4, p.~041304, 2015.

\bibitem{bayat2021review}
M.~Bayat, W.~Dong, J.~Thorborg, A.~C. To, and J.~H. Hattel, ``A review of
  multi-scale and multi-physics simulations of metal additive manufacturing
  processes with focus on modeling strategies,'' {\em Additive Manufacturing},
  vol.~47, p.~102278, 2021.

\bibitem{mehrpouya2022multimaterial}
M.~Mehrpouya, D.~Tuma, T.~Vaneker, M.~Afrasiabi, M.~Bambach, and I.~Gibson,
  ``Multimaterial powder bed fusion techniques,'' {\em Rapid prototyping
  journal}, 2022.

\bibitem{francois2017modeling}
M.~M. Francois, A.~Sun, W.~E. King, N.~J. Henson, D.~Tourret, C.~A. Bronkhorst,
  N.~N. Carlson, C.~K. Newman, T.~Haut, J.~Bakosi, {\em et~al.}, ``Modeling of
  additive manufacturing processes for metals: Challenges and opportunities,''
  {\em Current Opinion in Solid State and Materials Science}, vol.~21, no.~4,
  pp.~198--206, 2017.

\bibitem{yan2018modeling}
W.~Yan, S.~Lin, O.~L. Kafka, C.~Yu, Z.~Liu, Y.~Lian, S.~Wolff, J.~Cao, G.~J.
  Wagner, and W.~K. Liu, ``Modeling process-structure-property relationships
  for additive manufacturing,'' {\em Frontiers of Mechanical Engineering},
  vol.~13, pp.~482--492, 2018.

\bibitem{khairallah2016laser}
S.~A. Khairallah, A.~T. Anderson, A.~Rubenchik, and W.~E. King, ``Laser
  powder-bed fusion additive manufacturing: Physics of complex melt flow and
  formation mechanisms of pores, spatter, and denudation zones,'' {\em Acta
  Materialia}, vol.~108, pp.~36--45, 2016.

\bibitem{yu2022quantitative}
T.~Yu and J.~Zhao, ``Quantitative simulation of selective laser melting of
  metals enabled by new high-fidelity multiphase, multiphysics computational
  tool,'' {\em Computer Methods in Applied Mechanics and Engineering},
  vol.~399, p.~115422, 2022.

\bibitem{furstenau2020generating}
J.-P. F{\"u}rstenau, H.~Wessels, C.~Wei{\ss}enfels, and P.~Wriggers,
  ``Generating virtual process maps of slm using powder-scale sph
  simulations,'' {\em Computational Particle Mechanics}, vol.~7, no.~4,
  pp.~655--677, 2020.

\bibitem{afrasiabi2021multi}
M.~Afrasiabi, C.~L{\"u}thi, M.~Bambach, and K.~Wegener, ``Multi-resolution sph
  simulation of a laser powder bed fusion additive manufacturing process,''
  {\em Applied Sciences}, vol.~11, no.~7, p.~2962, 2021.

\bibitem{meier2021novel}
C.~Meier, S.~L. Fuchs, A.~J. Hart, and W.~A. Wall, ``A novel smoothed particle
  hydrodynamics formulation for thermo-capillary phase change problems with
  focus on metal additive manufacturing melt pool modeling,'' {\em Computer
  Methods in Applied Mechanics and Engineering}, vol.~381, p.~113812, 2021.

\bibitem{afrasiabi2022smoothed}
M.~Afrasiabi, C.~L{\"u}thi, M.~Bambach, and K.~Wegener, ``Smoothed particle
  hydrodynamics modeling of the multi-layer laser powder bed fusion process,''
  {\em Procedia CIRP}, vol.~107, pp.~276--282, 2022.

\bibitem{afrasiabi2022effect}
M.~Afrasiabi, D.~Keller, C.~L{\"u}thi, M.~Bambach, and K.~Wegener, ``Effect of
  process parameters on melt pool geometry in laser powder bed fusion of
  metals: A numerical investigation,'' {\em Procedia CIRP}, vol.~113,
  pp.~378--384, 2022.

\bibitem{dao2021simulations}
M.~H. Dao and J.~Lou, ``Simulations of laser assisted additive manufacturing by
  smoothed particle hydrodynamics,'' {\em Computer Methods in Applied Mechanics
  and Engineering}, vol.~373, p.~113491, 2021.

\bibitem{luethi2023adaptive}
C.~L{\"u}thi, M.~Afrasiabi, and M.~Bambach, ``An adaptive smoothed particle
  hydrodynamics (sph) scheme for efficient melt pool simulations in additive
  manufacturing,'' {\em Computers \& Mathematics with Applications}, vol.~139,
  pp.~7--27, 2023.

\bibitem{canelas2013generalized}
R.~Canelas, R.~Ferreira, A.~Crespo, and J.~Dom{\'\i}nguez, ``A generalized
  sph-dem discretization for the modelling of complex multiphasic free surface
  flows,'' in {\em Proceedings of the 8th International SPHERIC Workshop},
  pp.~74--79, Smooth Particle Hydrodynamics Research and Engineering
  International~…, 2013.

\bibitem{russell2018numerical}
M.~Russell, A.~Souto-Iglesias, and T.~Zohdi, ``Numerical simulation of laser
  fusion additive manufacturing processes using the sph method,'' {\em Computer
  Methods in Applied Mechanics and Engineering}, vol.~341, pp.~163--187, 2018.

\bibitem{khairallah2014mesoscopic}
S.~A. Khairallah and A.~Anderson, ``Mesoscopic simulation model of selective
  laser melting of stainless steel powder,'' {\em Journal of Materials
  Processing Technology}, vol.~214, no.~11, pp.~2627--2636, 2014.

\bibitem{liao2020fbstab}
D.~Liao-McPherson and I.~Kolmanovsky, ``Fbstab: A proximally stabilized
  semismooth algorithm for convex quadratic programming,'' {\em Automatica},
  vol.~113, p.~108801, 2020.

\bibitem{yadroitsev2010single}
I.~Yadroitsev, A.~Gusarov, I.~Yadroitsava, and I.~Smurov, ``Single track
  formation in selective laser melting of metal powders,'' {\em Journal of
  Materials Processing Technology}, vol.~210, no.~12, pp.~1624--1631, 2010.

\bibitem{trapp2017situ}
J.~Trapp, A.~M. Rubenchik, G.~Guss, and M.~J. Matthews, ``In situ absorptivity
  measurements of metallic powders during laser powder-bed fusion additive
  manufacturing,'' {\em Applied Materials Today}, vol.~9, pp.~341--349, 2017.

\end{thebibliography}

\end{document}